# Reference-Beam-Assisted Perturbation-Resistant Imaging Through A Single Multimode Fiber

*Bowen Sun[1]‡, Maohan Li[1]‡, Jingwei Luo[1], Zhuoren Wan[1], Zhaoyang Wen[1,2], Yuan Chen[1], Wending Huang[1], Zijian Wang[1,2]*, Ming Yan[1,2,3]*, and Heping Zeng[1,2,4]*

[1]State Key Laboratory of Precision Spectroscopy, and Hainan Institute, East China Normal University, Shanghai, China

[2]Chongqing Key Laboratory of Precision Optics, Chongqing Institute of East China Normal University, Chongqing 401120, China

[3]Collaborative Innovation Center of Extreme Optics, Shanxi University, Taiyuan, Shanxi 030006, People's Republic of China

[4]Jinan Institute of Quantum Technology, Jinan, Shandong 250101, China

**Abstract:**

Multimode fiber (MMF) imaging is highly sensitive to mechanical perturbations, which alter modal coupling and disrupt the speckle-to-target mapping. Here, we propose a reference-beam-assisted method for image reconstruction across multiple fiber configurations. An independent reference beam propagates through the same MMF as the imaging signal without target modulation, providing target-independent transmission-state information through its speckle patterns without requiring a separate configuration-sensing structure. A neural network jointly processes the signal and reference speckles to reconstruct target images. Experiments with MNIST and Fashion-MNIST across 10 fiber configurations demonstrate improved reconstruction quality, faster training convergence, and more accurate configuration identification. Further evaluation across 40 configurations confirms the reconstruction benefit of reference-beam assistance over a larger set of perturbed fiber states. These results provide a foundation for developing perturbation-adaptive MMF imaging systems for flexible endoscopy and minimally invasive optical diagnostics.

## Introduction

Multimode fibers (MMFs) support a large number of guided modes within a small cross-sectional area, providing a high capacity for spatial information transmission. They have therefore attracted considerable interest in applications such as minimally invasive endoscopic imaging [1–3], high-speed fiber imaging [4], three-dimensional imaging [5–6], and optical fiber communications [7–12]. However, modal propagation and coupling in an MMF are highly sensitive to the fiber configuration. Mechanical deformations such as bending and twisting alter the relative phases and coupling among the guided modes, resulting in substantial changes in both the output speckle pattern and the input–output mapping [13–15]. Maintaining reliable image reconstruction under perturbations remains a central challenge for MMF imaging.

MMF imaging generally relies on wavefront shaping [16–17], optical phase conjugation [18], or transmission-matrix measurements [19–22] to characterize or compensate for the complex transmission of optical fields. These approaches achieve high-quality focusing and image reconstruction when the fiber remains stable, but require recalibration when configuration changes invalidate the original transmission response. To improve adaptability to perturbations, compressive transmission-matrix measurements [23], binary silicon-photonic phased-array modulation [24], and digital micromirror device (DMD)-based full-field wavefront shaping [25] have been developed to increase calibration efficiency and enable dynamic optical-field control. However, these approaches still rely on updated calibration or feedback measurements to maintain compensation as the fiber configuration changes, adding measurement and control requirements to the imaging system.

Deep learning provides a data-driven approach to dynamic MMF imaging [26–28]. Neural networks directly learn the mapping from MMF output speckle patterns to target images [29–33], allowing image reconstruction without explicitly determining the complete transmission matrix. Joint training across multiple perturbation states [34], variational autoencoding [35], and transfer learning [36] improve adaptation to changes in fiber configuration. More recently, self-supervised dynamic updating has enabled online adaptation to evolving transmission conditions, improving long-term imaging stability [37]. Nevertheless, most existing data-driven MMF imaging methods still lack an independent measurement of the fiber state and rely primarily on the output speckle pattern for target reconstruction. To provide explicit state information, Zhu et al. integrated an MMF with a fiber Bragg grating (FBG) array and supplied the measured configuration parameters as additional network inputs, improving image reconstruction under perturbations [38]. However, this strategy requires an additional FBG sensing structure and its associated readout, increasing the complexity of the imaging system.

In this work, we develop a reference-beam-assisted framework for image reconstruction through an MMF across multiple fiber configurations. By propagating alongside the imaging signal in the same fiber, an independent reference beam captures configuration-dependent changes in optical transmission without a separate sensing path. The resulting reference speckles complement the signal speckles as inputs to a neural network, enabling image recovery together with configuration recognition. We validate this approach using MNIST and Fashion-MNIST targets under different fiber configurations. Incorporating the reference information enhances reconstruction fidelity, facilitates network convergence, and improves the accuracy of fiber configuration identification, supporting its utility in multi-configuration MMF imaging.

**Basic principle and experimental setup**

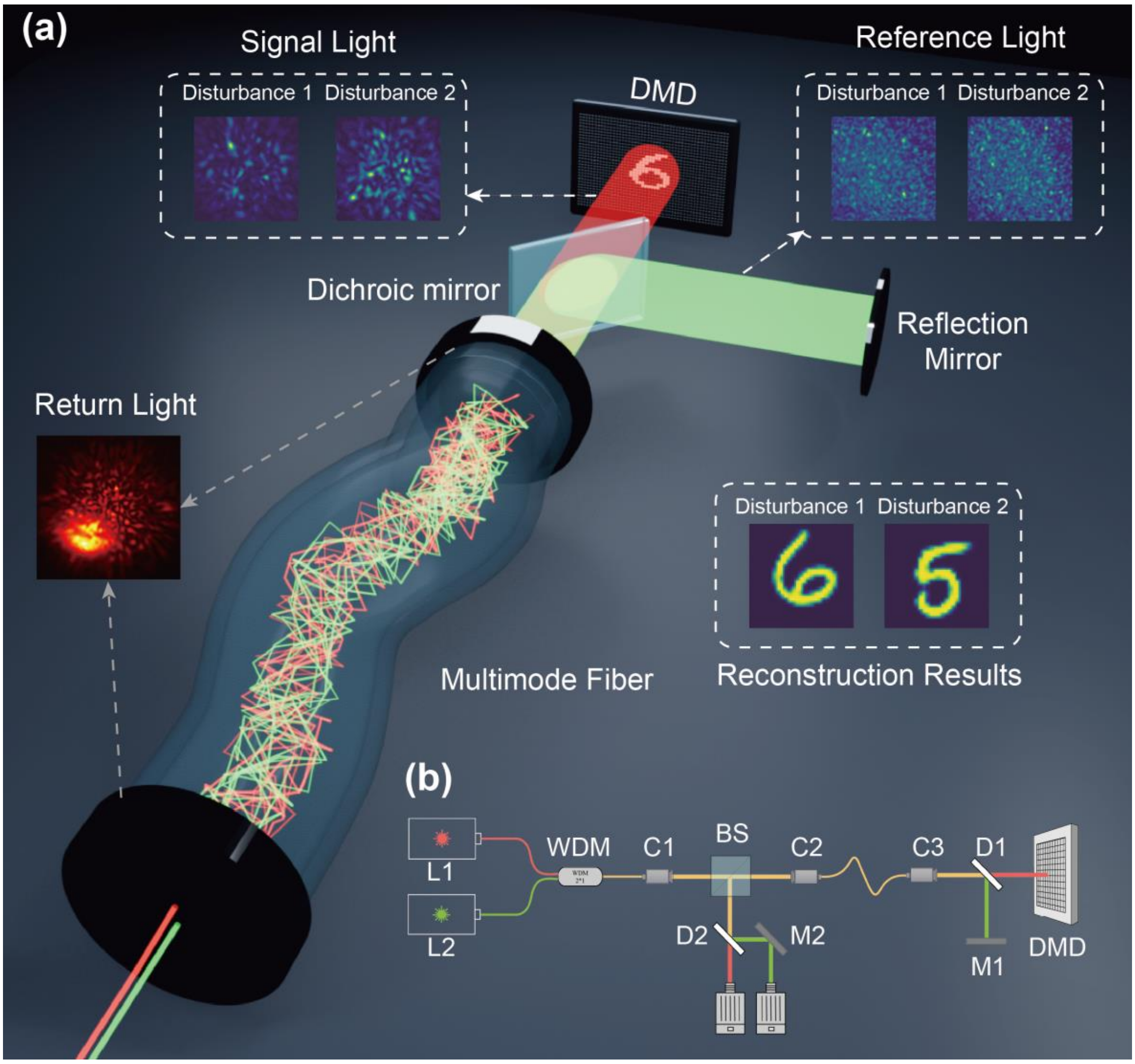

**Figure 1 | Schematic of the experimental setup. a,** Schematic of reference-beam-assisted imaging through a dynamically perturbed multimode fiber. The signal and reference beams co-propagate through the multimode fiber (MMF) and produce speckle responses corresponding to the fiber configuration. At the distal end of the fiber, the signal beam is modulated by the DMD and then returns whereas the reference beam is directly reflected back by a mirror. **b,** Schematic of the experimental optical setup. L1 and L2 denote the signal and reference light sources, respectively; WDM, wavelength-division multiplexer; C1-C3, collimators; BS, beam splitter; D1 and D2, dichroic mirrors; M1 and M2, mirrors; and DMD, digital micromirror device.

As illustrated in Figure 1(a), the system employs a round-trip transmission architecture based on a single MMF. A signal beam for imaging and a visible reference beam for probing the fiber transmission state are jointly coupled into the proximal end of the MMF.

Both beams form complex speckle fields through modal interference, which are highly sensitive to changes in the fiber configuration. At the distal end, a dichroic mirror separates the two beams according to wavelength. The signal beam is modulated with the target pattern by a digital micromirror device (DMD) and then returns through the MMF, whereas the reference beam is reflected back by a mirror without target modulation. The signal and reference speckles acquired under different fiber configurations generated using a manual translation stage are jointly processed by a neural network for target image reconstruction and fiber configuration identification. Because of the round-trip geometry, back-reflections from the MMF end facets and other optical interfaces are inevitably introduced. Some of these reflections, such as those from the distal collimator C3, may carry configuration-dependent information but contain no controlled target modulation; they are treated as noise in the present experiments.

The experimental setup is shown in Figure 1(b). The signal beam is generated by a mode-locked laser L1 with a center wavelength of 1550 nm, while the reference beam is provided by a visible laser L2 operating at a center wavelength of 532 nm. The two beams are combined by a wavelength division multiplexer and directed through collimator C1 into a 1-m-long MMF with a 200-μm core diameter and a numerical aperture of 0.22 (LBTEK, MMC200/400L-0.22-PC-1). At the distal end of the fiber, the output is collimated by C3 and spectrally separated by dichroic mirror D1 (Thorlabs, DMLP950). The signal beam is modulated by the DMD (CASMicrostar, DMD-1K065-03-16) and then retraces the original optical path. The reference beam enters a reference arm and is directly reflected back without being modulated. At the proximal end, the returning light is separated from the incident path by beam splitter BS and is then spectrally divided by dichroic mirror D2 into the signal and reference channels. The corresponding speckle intensity distributions are recorded with an infrared camera (Daheng Imaging, MARS-138-95GM-P-TN-SWIR) and a visible light camera (Hikrobot, MV-CA016-10UM), respectively. During the experiment, the middle section of the MMF is mounted on a manual linear translation stage with a travel range of 100 mm and a precision of 0.05 mm, allowing the fiber bending to be varied by adjusting the transverse displacement.

**Network architecture**

Reference speckles provide fiber-state information to assist image reconstruction under perturbations. We therefore jointly process signal and reference speckles using a hybrid attention-gated U-Net (HAG-U-Net) to reconstruct target images and identify fiber configurations, as shown in Figure 2. The speckle pairs form a $2 \times 128 \times 128$ input tensor, which is mapped to 10 feature channels by a $1 \times 1$ convolution followed by Leaky ReLU. Each HAG module sequentially applies channel and spatial attention to encoder features before concatenation with decoder features [Figure 2(b)–(d)]. Channel weights are derived from joint encoder and decoder features using global average and max pooling. Spatial weights are then generated from the channel-refined encoder features and decoder features. The decoder produces a $10 \times 64 \times 64$ output through a final $1 \times 1$ convolution followed by sigmoid activation.

For each fiber configuration, 3,000 signal speckle images, one background image, and one reference speckle image are acquired. After background subtraction from the signal images, the signal and reference speckles are cropped, resized to $128 \times 128$ pixels using bilinear interpolation, and normalized. Target patterns are resized to $64 \times 64$ pixels and assigned to the output channel corresponding to the fiber configuration, with all other channels set to zero. Samples from each configuration are split into training, validation, and test sets in proportions of 84%, 11%, and 5%, respectively.

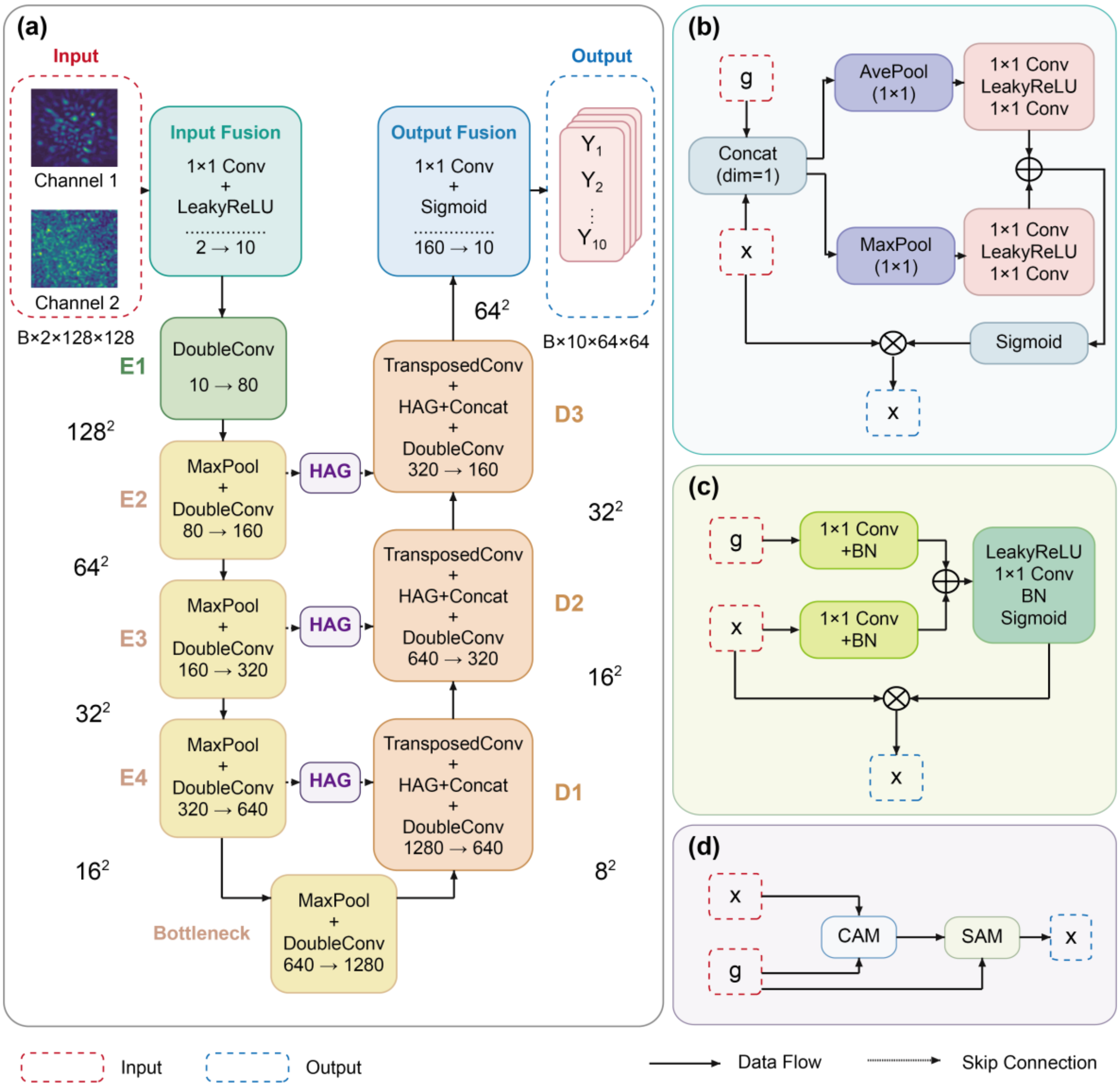

**Figure 2 | Hybrid Attention-Gated U-Net architecture. a**, Overall architecture of HAG-U-Net. **b**, Channel attention module (CAM). **c**, Spatial attention module (SAM). **d**, Hybrid attention-gating (HAG) module, where x denotes the encoder feature and g denotes the decoder feature.

For network training, the model is implemented in Python using the PyTorch deep-learning framework and trained on an NVIDIA GeForce RTX 5080 GPU. Mean squared error (MSE) is used as the loss function and is calculated over all output channels. The network parameters are optimized with AdamW. The initial learning rate is set to $10^{-3}$ and is multiplied by 0.5 every five epochs. Dropout regularization is applied with a dropout probability of 0.25 to mitigate overfitting. The batch size is set to 32. The model checkpoint yielding the lowest validation loss is retained as the final model.

## Results

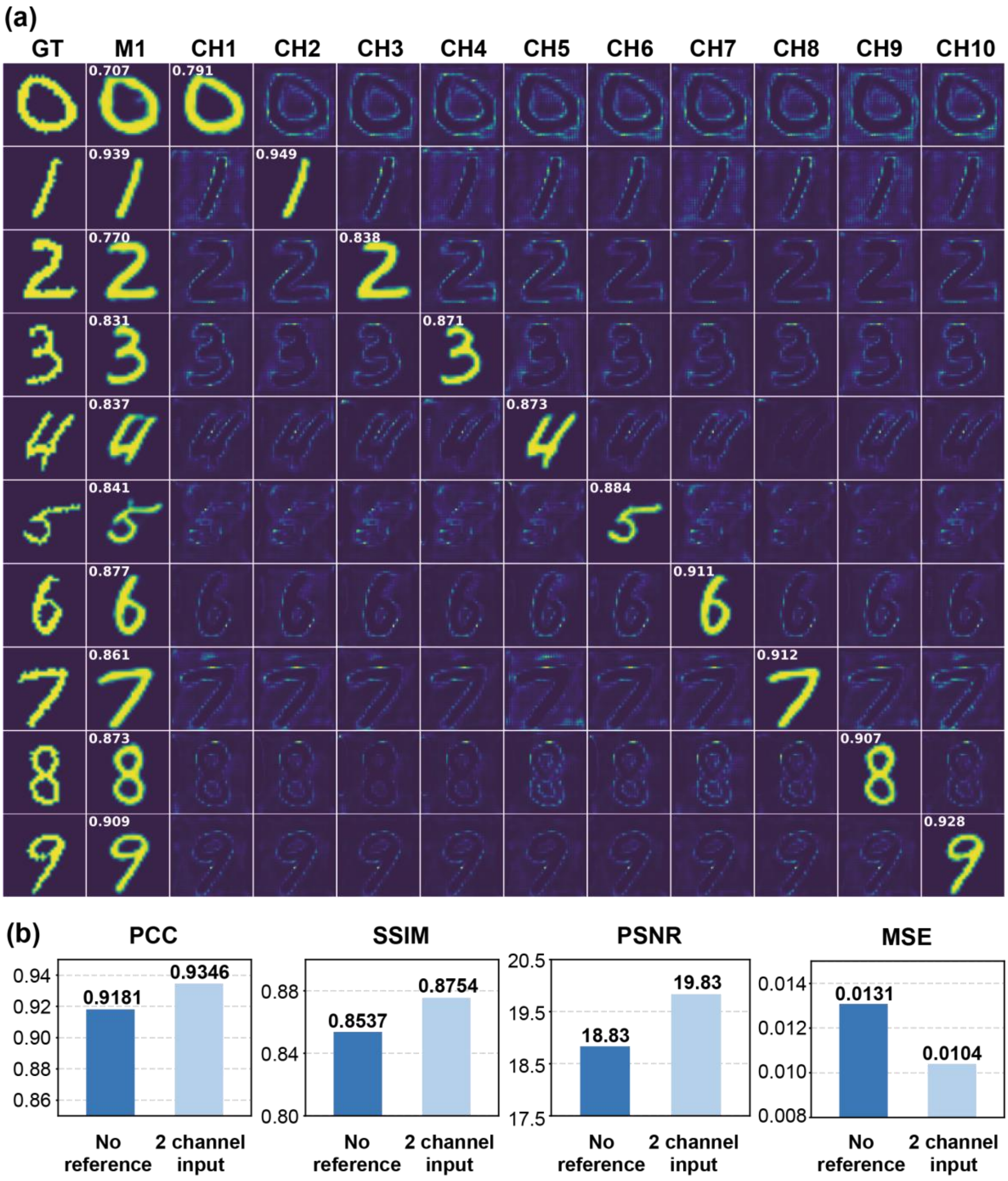

**Figure 3 | Comparison of image reconstruction performance across 10 fiber configurations with and without reference-beam assistance. a**, Representative reconstructions of MNIST targets under 10 different fiber configurations. GT, ground-truth target; M1, the method without the reference beam; CH1–CH10, the outputs of the reference-assisted model in the channels corresponding to the different fiber configurations. The annotated values are the structural similarity index (SSIM) between each reconstruction and the ground truth. **b**, Comparison of the average reconstruction performance of the two models on the test set, evaluated using the Pearson correlation

coefficient (PCC), structural similarity index (SSIM), peak signal-to-noise ratio (PSNR), and mean squared error (MSE).

To evaluate image reconstruction across multiple fiber configurations, we generated 10 configurations by shifting the slider of the manual translation stage in 5-mm increments. Two models were compared: a model using only the signal speckle as input and a reference-assisted model jointly using the signal and reference speckle patterns. Figure 3(a) presents representative reconstructions of MNIST targets under the 10 configurations. For the reference-assisted model, CH1–CH10 denote the output channels assigned to the corresponding fiber configurations. The quantitative comparison in Figure 3(b) shows improvements in PCC, SSIM, PSNR, and MSE with reference-beam assistance. These results confirm that the reference speckles provide the network with effective information about the fiber state, thereby improving image-reconstruction quality under dynamical conditions.

To assess the extent of the transmission changes across the fiber configurations used in Figure 3, we calculated the PCC between each measured speckle pattern and its corresponding baseline pattern. For both the signal and reference beams, the correlations between speckle patterns corresponding to different perturbation states decrease rapidly and remain low overall, indicating substantial changes in the transmission response (Supplementary Figure S1). Such changes complicate image reconstruction using a mapping learned under a fixed configuration and highlight the value of incorporating fiber-state information when reconstructing images across multiple configurations.

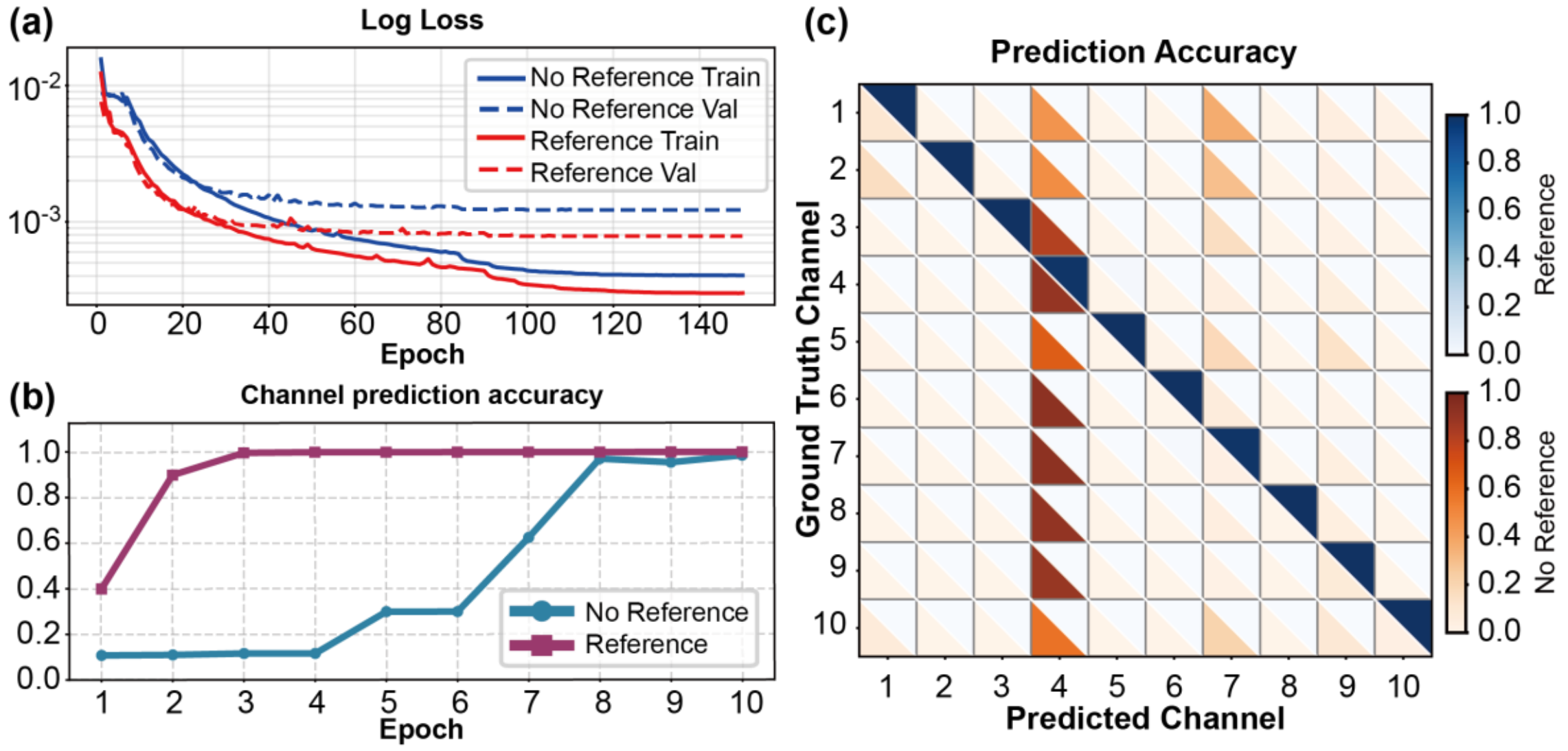

**Figure 4 | Comparison of network training and fiber configuration identification with and without reference-beam assistance. a**, Training and validation losses of the two methods during training. b, Fiber configuration classification accuracy of the models with and without the reference beam over the first 10 training epochs. **c**, Confusion matrices for classifying 10 fiber configurations at the third training epoch. Within each

cell, the blue and orange regions represent the predictions of the reference-assisted and no-reference methods, respectively.

In addition to improving the final reconstruction performance, the reference channel has a pronounced effect on the network training process. Figure 4(a) shows the training and validation losses of the models with and without the reference beam. With the reference beam, both losses decrease more rapidly and ultimately converge to lower values. This result indicates that the transmission-state information carried by the reference speckle reduces the difficulty of learning the dynamic speckle-to-target mapping.

The benefit of the state information provided by the reference beam is even more evident in the fiber configuration classification task. The channel with the highest mean intensity is selected as the predicted fiber configuration class. Figure 4(b) shows the classification accuracy of the two methods over the first 10 training epochs. With the reference beam, the classification accuracy increases rapidly from the early stage of training and reaches a high, stable level within only a few epochs. Figure 4(c) evaluates the models at the third training epoch and shows the confusion matrices for classifying the 10 fiber configurations. The reference speckle patterns from different configurations are highly decorrelated and thus serve as distinctive “fingerprints” of the corresponding fiber states, enabling the network to rapidly associate each reference speckle pattern with its fiber configuration.

To further evaluate the adaptability of the reference-assisted strategy to different target types, we extend the experiments to the Fashion-MNIST dataset. Figure 5(a) shows representative reconstructions obtained with and without reference-beam assistance under 10 fiber configurations. Supplementary Figure S2 further provides a quantitative comparison of the reconstruction performance of the two methods on Fashion-MNIST. The classification accuracies of the two models over the first 10 training epochs are presented in Figure 5(b). These results show that the transmission-state information carried by the reference beam is not specific to a particular target class and remains applicable to target images with more complex structures and greater inter-class variation.

The robustness of the reference-assisted approach was further assessed under 40 fiber configurations, generated by shifting the slider in 2.5-mm increments to deform the MMF. Figure 5(c) shows the training, validation, and test losses for the two approaches over 150 training epochs. As training proceeds, both models exhibit decreasing training and validation losses, indicating that the networks progressively fit the training data containing all 40 fiber configurations. However, the test loss reaches a minimum at approximately 20 epochs and then gradually increases with further training, indicating a risk of overfitting at later stages. This behavior may be associated with the limited number of samples and the increased complexity introduced by the larger number of perturbation states. Next, Figure 5(d) more directly illustrates the difference in image reconstruction performance with and without reference light across 40 perturbation states. The reference-assisted model maintains comparatively stable and clear reconstructions, whereas the no-reference method occasionally exhibits pronounced blurring or structural distortion. These results demonstrate the effectiveness of the reference beam in complex multi-perturbation environments even with a limited number of training epochs. They also indicate that appropriate regularization strategies should be introduced in future work to alleviate overfitting and further improve model generalization.

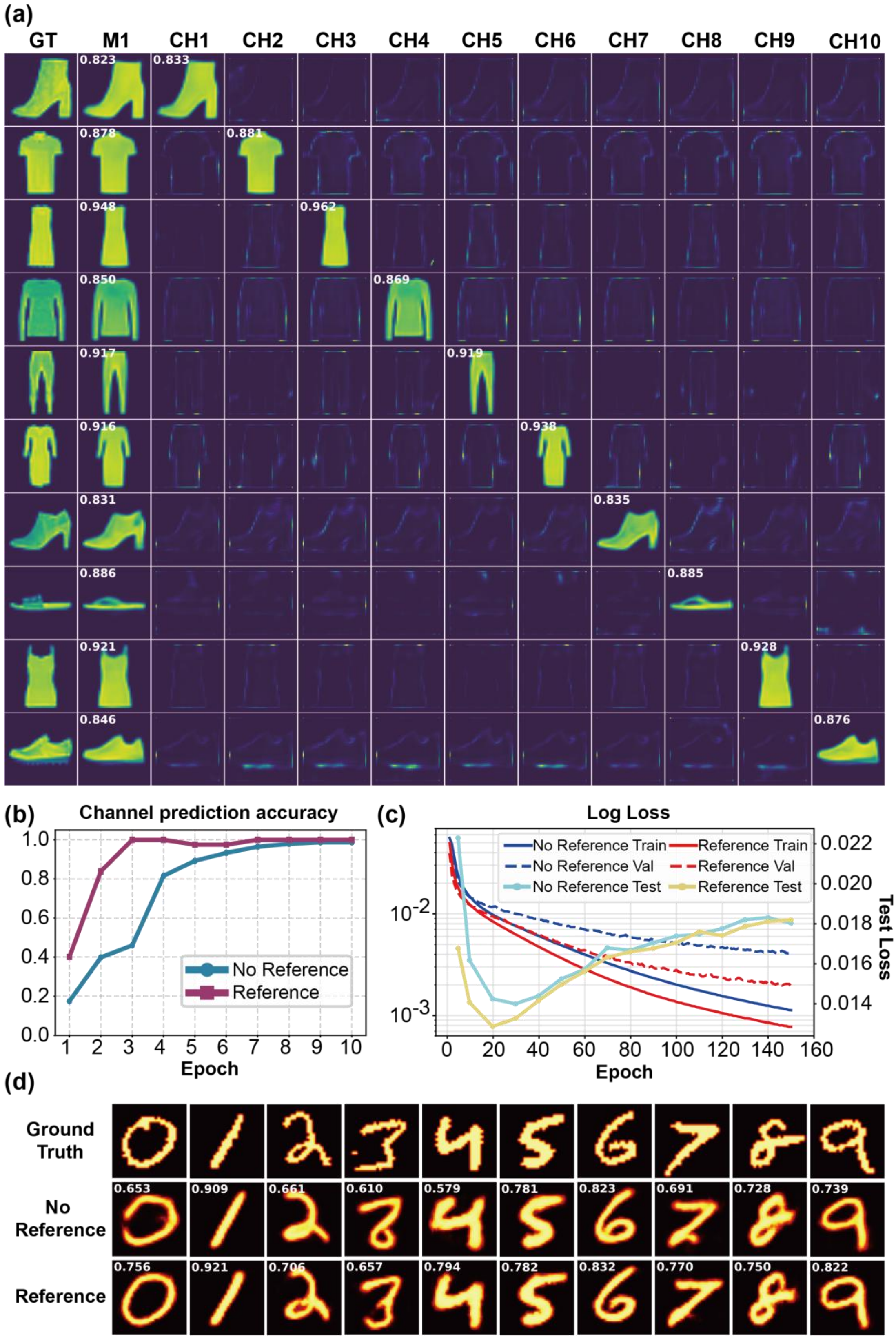

**Figure 5 | Reconstruction and fiber configuration classification on Fashion-MNIST and reconstruction performance on MNIST. a**, Representative reconstructions of Fashion-MNIST targets under 10 different fiber configurations. GT, ground-truth target; M1, the method without the reference beam; CH1–CH10, the outputs of the reference-assisted model in the channels corresponding to the different fiber configurations. The annotated values are the structural similarity index (SSIM) between each reconstruction and the ground truth. **b**, Classification accuracy on the Fashion-MNIST dataset during the early stage of training for the methods with and without the reference beam. **c**, Training, validation, and test losses of the two methods under 40 fiber configurations; the training and validation loss curves are plotted on a logarithmic vertical scale. **d**, Representative MNIST test reconstructions by the two methods under 40 fiber configurations. The annotated values are the SSIM between each reconstructed image and its corresponding ground truth.

## Discussion

Our work introduces a reference-beam-assisted framework that incorporates fiber transmission-state information into image reconstruction through a single MMF. Joint processing of signal and reference speckles with HAG-U-Net improves reconstruction quality and facilitates configuration identification across the tested fiber configurations. The adaptability of the framework could be further improved through broader configuration sampling, transfer learning [36], and self-supervised dynamic updating [37]. The present training and testing involve discrete configurations generated by a manual translation stage; combining more extensive sampling with model adaptation could help extend the method toward unseen configurations and continuously varying perturbations. In addition, weak target contours are observed in non-corresponding output channels, and edge-extraction and network tests suggest that they arise from cross-channel feature sharing rather than inherent edge robustness to perturbations. The increased leakage under overfitting further motivates strategies that suppress unintended responses while preserving useful shared features.

The framework also offers opportunities for further development in system design and imaging functionality. Preliminary experiments show that back-reflected light can also assist image reconstruction (Supplementary Figure S3), suggesting a simpler implementation using existing reflections. Reconstructions of spatial light modulator (SLM)-encoded facial phase patterns further suggest the potential for phase imaging without an additional interferometric arm (Supplementary Figure S4). Furthermore, optical frequency combs, which provide multiple spectral channels for spectroscopy, time-domain imaging, and ranging [39–45], offer another route to extending the framework. Individual comb lines could serve as wavelength-resolved references, providing complementary information about the fiber transmission state. Drawing on recent advances in hyperspectral dual-comb imaging [46], this approach could be further developed for self-referenced multispectral imaging.

## Corresponding Authors

Zijian Wang; Email: zjwang@lps.ecnu.edu.cn

Ming Yan; Email: myan@lps.ecnu.edu.cn

## Abbreviations

MMF, multimode fiber; HAG-U-Net, Hybrid Attention-Gated U-Net; HAG, hybrid attention-gating; DMD, digital micromirror device; FBG, fiber Bragg grating; WDM, wavelength-division multiplexer; PCC, Pearson correlation coefficient; SSIM, structural similarity index; PSNR, peak signal-to-noise ratio; MSE, mean squared error; GPU, graphics processing unit; SLM, spatial light modulator.